\documentclass[12pt,a4paper,english,superscriptaddress,aps,preprint]{revtex4}
\usepackage{amsmath}
\usepackage{amssymb}
\usepackage{graphicx}
\usepackage{hhline}
\usepackage{mwe}
\usepackage{amsbsy}
\usepackage{textcomp}
\usepackage{commath}

\usepackage{stackengine}
\stackMath

\makeatletter
\usepackage{babel}
\newcommand{\bea}{\begin{eqnarray}}
\newcommand{\eea}{\end{eqnarray}}

\newcommand{\pa}{\partial}

\newcommand{\be}{\begin{equation}}
\newcommand{\ee}{\end{equation}}
\numberwithin{equation}{section}

\begin{document}
\immediate\write16{<<WARNING: LINEDRAW macros work with emTeX-dvivers
                    and other drivers supporting emTeX \special's
                    (dviscr, dvihplj, dvidot, dvips, dviwin, etc.) >>}

\title{  BPS deformations of the Skyrme model}
\date{29-11-2020}
\author{J. M. Queiruga}
\email{josemanuel.fernandezq@ehu.eus}
\affiliation{Department of Theoretical Physics, UPV/EHU,
48080, Bilbao, Spain}

\begin{abstract}
We study several deformations of the Skyrme model in three dimensions with self-dual sectors of arbitrary baryonic charge. We show that, for a family of background metrics as well as for a family of field dependent couplings, the model has one BPS sector, which may have any topological charge. We also study the gravitating case, where there are infinite BPS sectors, provided that a cosmological constant is added to the model.

  \end{abstract}

\maketitle

\section{Introduction}

Although the details still remain to be fully understood, the Skyrme model \cite{Skyrme1,Skyrme2} can be interpreted as a low energy limit of QCD when the number of quark colors tends to infinity \cite{witten1, witten2}. In particular, it can be used to describe some properties of nuclei, like for example binding energies. In this effective approach the fundamental fields are interpreted as pions, while the topological solutions of the model, called Skyrmions, represent nuclei. From a mathematical point of view, a Skyrmion is a map from a three dimensional manifold (a one point compactification of $\mathbb{R}^3$ in the standard formulation) to $SU(2)$. Since $\pi_3(SU(2))=\mathbb{Z}$ the solutions are classified by a topological number, normally identified with the baryonic charge, and therefore they are topologically stable. 

Despite of its rich mathematical structure, the original formulation of the Skyrme model only provides a rough description of some properties of the low energy QCD \cite{Nappi}. For example, regarding the applications to nuclear physics, one of the main shortcomings of the model is that the classical binding energies predicted for nuclei are too big. One possible strategy to reduce the binding energies is to look for modifications of the model that posses self-dual sectors. There have been several attempts in the literature to confront this problem either by adding higher-derivative terms \cite{adam-BPS-Skyrme}, or by adding extra fields \cite{sutcliffe2, sutcliffe3}. 

In this work we will explore two extra possibilities. The first one is the modification of the base space manifold. It is known that, the charge one Skyrmion cannot be BPS if the space manifold is $\mathbb{R}^3$. This is because the hypothetical BPS solution would be an isometry between the base and target space manifolds and there are not isometries between $\mathbb{R}^3$ and $\mathbb{S}^3$. If one considers the formulation of the model in $\mathbb{S}^3$, the charge one Skyrmion is the identity map between three-spheres, which is trivially an isometry and it is automatically BPS \cite{Manton2}. The second possibility is to allow the coupling constants to be ``running couplings" or dielectric terms \cite{Adam1}. This allows again for charge one BPS Skyrmions. In the present work we generalize these results to higher-charge Skyrmions.

We will also study a gravitating Skyrme model. It is remarkable that, the gravitating $B=1$ case it is actually solvable \cite{Canfora1,Canfora2} and self-dual in the matter sector, provided that a cosmological constant is added to the model. We will also extend this result to higher-charges.

This paper is organized as follows. In Sec. \ref{curved-space} we describe a Skyrme model with BPS sectors in curved space. In Sec. \ref{dielectric}, we analyze several dielectric Skyrme models with arbitrary BPS sectors. In Sec. \ref{gravitating} we study the gravitating Skyrme model with cosmological constant. Finally, Sec. \ref{conclusions} contains our conclusions and further discussion.


\section{The Skyrme model in curved spacetime}
\label{curved-space}

 In this paper we will focus on static configurations, therefore we will mainly be interested in three dimensional manifolds (spatial slices at fixed time of the four dimensional spacetime manifold). The original Skyrme model consists of two terms, a Dirichlet type term, quadratic in derivatives, and a Skyrme term, quartic in derivatives. The static energy functional over an arbitrary three-dimensional manifold  $\mathcal{M}_3$, characterized by a metric $g_{ij}$, has the following form
 \be
E=\int d^3 x\sqrt{-g} \left( -\frac{f^2}{2}g^{ij}\text{tr} \, R_i R_j-\frac{1}{16 e^2}g^{i j}g^{kl}\text{tr}\lbrack R_i,R_k\rbrack \lbrack R_j,R_l\rbrack\right),
 \ee
 where  $R_i=\pa_i U U^{-1}$ are the $su(2)$-valued currents, $U\in SU(2)$ contains the fundamental degrees of freedom and $e,f$ are constants. It is very useful to define the strain tensor $D_{ij}$ as follows
 \be\label{strain}
 D_{ij}=-\frac{1}{2} \text{tr}R_i R_j.
 \ee
 By using the eigenvalues of $D_{ij}$ the static energy functional can be cast as
 \be\label{energy}
 E=\int d^3 x\sqrt{-g} \left(f \omega_m \pm \frac{1}{2e }\vert \epsilon_{mnl}\vert\omega_n \omega_l \right)^2\mp \frac{f}{e}6\int d^3x\lambda_1 \lambda_2\lambda_3,
 \ee
 where $\omega_m=e^{\,i}_m \lambda_i$, $e^{i}_m$ is the dreibein verifying  $g^{ij}=e^i_{m}e^{ j}_m$. The energy functional (\ref{energy}) gives a lower bound
 \be\label{bound}
\vert E \vert \geq \frac{f}{e} 6\int d^3x \lambda_1 \lambda_2\lambda_3,
 \ee
 which is saturated when the first term in (\ref{energy}) vanishes. This gives the following set of BPS equations 
\be\label{BPS}
f \omega_m \pm \frac{1}{2e}\vert \epsilon_{mnl}\vert\omega_n \omega_l=0.
\ee

Of course, in general it is not guaranteed that equations (\ref{BPS}) have nontrivial solutions. For example, as we have said, in $\mathbb{R}^3$, and for $e,f$ constants, (\ref{BPS}) are only verified for constant fields. 


\subsection{BPS structure}

We will analyze under which circumstances the equations (\ref{BPS}) can be satisfied. Since we are interested in looking for a single positive answer to the problem, we will restrict the possible solutions to only those that respect a particular form. As we will see later, this choice not only works as a pro-example, but also allows us to find trivial analytical solutions on each topological sector.  Let us consider, then,  the following  unit vector
\be
\hat{\bold{n}}_R=\frac{1}{1+\vert R\vert^2}\left(2\,\text{Re}\, R,2\,\text{Im}\, R,1-\vert R\vert^2 \right),
\ee
where $R$ is a rational function of $z$. We define the $SU(2)$ Skyrme field in the rational map ansatz (\cite{Manton1}, \cite{Manton3}) as follows
\be\label{ansatz}
U(r,z)=e^{i \xi(r) \hat{\bold{n}}_R \cdot \bar{\bold{\sigma} }},
\ee
where $ \bar{\bold{\sigma} }=(\sigma_1,\sigma_2,\sigma_3)$ are the Pauli matrices. The eigenvalues of the strain tensor (\ref{strain}) for the ansatz (\ref{ansatz}) are given by
\be\label{eigen1}
\lambda_1^2=\xi'^2(r),\,\,\, \lambda_2^2= \lambda_3^2=\frac{\sin^2 \xi(r)}{r^2}\frac{\left(1+\vert z\vert^2\right)^2}{\left(1+\vert R\vert^2\right)^2}\abs{ \frac{dR}{dz}}^2.
\ee

Taking into account the form of the eigenvalues of the strain tensor given in (\ref{eigen1}) and the general form of the (possible) BPS equations, we consider the following base space metric
\be\label{metric}
g_{ij}=\frac{4 r_0^2}{\left(1+r^2\right)^2}\text{diag} \left(1,r^2\left(\frac{1+\vert z\vert^2}{1+\vert Q\vert^2}\right)^2\abs{ \frac{dQ}{dz}}^2,r^2\left(\frac{1+\vert z\vert^2}{1+\vert Q\vert^2}\right)^2\abs{ \frac{dQ}{dz}}^2 \sin^2\theta \right).
\ee
where $z=\tan\left(\frac{\theta}{2}\right)e^{i\phi}$ and $Q$ is a fixed rational function of $z$. Note that for $Q(z)=z$, and after the change of coordinates $r=\tan\psi/2$, the metric (\ref{metric}) takes the form of the round metric in $\mathbb{S}^3$ of radius $r_0$
\be
ds^2=r_0^2\left(d\psi^2+\sin^2\psi (d\theta^2+\sin^2\theta d\phi^2)\right).
\ee

 The BPS equations (\ref{BPS}) with the metric (\ref{metric}) take the form
\bea\label{eqBPS1}
ef \xi'(r)+ \frac{\left(1+\vert Q\vert^2\right)^2}{\left(1+\vert R\vert^2\right)^2}\frac{\vert R'\vert^2}{\vert Q'\vert^2}\frac{1+r^2}{ r_0 r^2}\sin^2(\xi(r))&=&0\\ \label{eqBPS2}
 \frac{1+\vert Q\vert^2}{1+\vert R\vert^2}\frac{\vert R' \vert}{\vert Q' \vert}\left(\frac{1}{r_0} \xi'(r)+2 \frac{ef}{1+r^2}\right)&=&0.
\eea
 If $ef=1/r_0$ these equations have a common solution
\be\label{solBPS}
\xi(r)=-2\arctan r+\pi, \quad R(z)=Q(z),
\ee
which interpolates between $\pi$ and $0$ for $r\in(0,\infty)$. In terms of the coordinate $\psi\in [0,\pi]$ the solution is 
\be
\xi(\psi)=\pi-\psi,
\ee
which represents the identity map. This solution was previously found for $B=1$ in $\mathbb{S}^3$ in \cite{Manton2} and in terms of the dielectric functions in \cite{Adam1} (in the latter, this solution corresponds to the choice $\alpha=1/2$ of that paper). The topological charge of the solution is actually determined by the choice of $R(z)$ vie the usual relation
\be
B=-\frac{1}{2\pi^2}\int \frac{\xi'(r)\sin^2 \xi(r)}{\left(1+\vert r\vert^2\right)^2}\abs{ \frac{dR}{dz}}^2 2i d\bar{z}dzdr=N,
\ee
where $N$ is the degree of the map $R$. As a consequence, not only the $B=1$ sector can be made BPS by choosing the round $\mathbb{S}^3$ to be the base space manifold, but any choice of (\ref{metric}) with $\text{deg}Q=N$ brings the $B=N$ sector to self-duality. Moreover, this BPS solution is given by the rational map ansatz defined by the same rational function defining the metric. This is behind the choice of the metric (\ref{metric}), which is the unique metric that allows for exact BPS solutions in the form of the rational map ansatz. In addition, it is straightforward to prove that (\ref{eqBPS1}) and (\ref{eqBPS2}) imply the Euler-Lagrange equations. Another feature of the solutions (\ref{solBPS}) is that the length scale determined by the Syrmion solutions, $L=1/ef$ must equal the radius $r_0$ of the base space manifold is order for the solutions to be BPS. It is worth noting that for any choice of metric on the form (\ref{metric}) the radial profile of the solutions in the self-dual sector is of the form given in (\ref{solBPS}), i.e. it is an identity map in the $\psi$ coordinate regardless of the topological charge, while the angular part is given by $Q(z)$. In principle, there may be other metric choices, not related to the rational map ansatz providing BPS sectors as well. The strategy for finding them follows the lines described above: fix another field ansatz, compute the eigenvalues for the strain tensor and choose $g_{ij}$ such that the BPS equations as satisfied simultaneously.


\section{Dielectric Skyrmions with higher-charge BPS sectors}
\label{dielectric}

We have considered so far, how modifications of the space manifold may generate certain BPS sectors (of any baryonic charge). That is, a judicious modification of the space metric in which the Skyrme model is formulated allows for first order equations saturating a Bogomolny bound. We discuss now if some ``mild" modification of the Skyrme model may lead to similar results, but now keeping the base space manifold to be $\mathbb{R}^3$. One possibility is the addition of impurities (background space-dependent fields). This was already considered in \cite{Naya} (see also \cite{tong, BPS-imp-susy, solvable-imp, vortex-imp-2, vortex-imp-3} for other BPS models with imourities), but we will not pursue this line. In \cite{Adam1}, the authors consider a dielectric Skyrme model, that consists of the original Skyrme model (quadratic plus quartic terms), multiplied by some dielectric functions. These dielectric functions depend on the field variables and therefore, they can be viewed as a sort of modification of the target space metric, even though these dielectric functions allow in general for more freedom than target-space metric modifications. They found that, with such a modification, it is possible to bring the $B=1$ Skyrmion to the BPS sector. We will show here that it is indeed possible to generate BPS sectors of arbitrary charge (but only one at a time) by modifying properly the dielectric functions. Let us consider the following action
 \be\label{energy-target}
E=\int d^3 x\left( -\frac{1}{2}G_t^{ij}(\pi^a)\text{tr} \, R_i R_j-\frac{1}{16 }G_t^{i j}(\pi^a)G_t^{kl}(\pi^a)\text{tr}\lbrack R_i,R_j\rbrack \lbrack R_k,R_l\rbrack\right),
 \ee
 where $G_t^{ij}(\pi^a)$ are functions of the target space variables, generically denoted by $\pi^a$, defined as follows
 \be
 G_t^{ij}(\pi^a)=\text{diag}(f_1^2(\pi^a),f_2^2(\pi^a),f_3^2(\pi^a)).
 \ee
 Note that we have reabsorbed the constant couplings $e,f$ in the definitions of the functions $f_i$. By completing the square, the energy (\ref{energy-target}) we can be expanded as follows
\bea
E&=&\int_{\mathbb{R}^3} \left(f_1 \lambda_1\pm f_2 f_3\lambda_2 \lambda_3\right)^2+ \left(f_2 \lambda_2\pm f_1 f_3\lambda_1  \lambda_3\right)^2+ \left(f_3 \lambda_3\pm f_1 f_2\lambda_1  \lambda_2\right)^2 \nonumber \\
 &&\mp 6 \int_{\mathbb{R}^3} f_1f_2f_3\lambda_1 \lambda_2 \lambda_3,
\eea
therefore
\be\label{target1}
\vert E \vert \geq 6 \int_{\mathbb{R}^3} f_1f_2f_3\lambda_1 \lambda_2 \lambda_3.
\ee
The bound is saturated if and only if
\be\label{BPSrel}
(f_1\lambda_1)^2=(f_2\lambda_2)^2=(f_3\lambda_3)^2=1.
\ee

It is well-known that, in the standard case this bound cannot be attained for any nontrivial configuration and, as we have said, it is also known that there is a choice of dielectric functions \cite{Adam1} that makes the $B=1$ sector BPS. Let us explore the rest of sectors. For the sake of concreteness let us take the rational map ansatz (\ref{ansatz}) whose target space coordinates $\pi^a$ are expanded in terms of  $\xi(r)$ and $R(z)$. The relations (\ref{BPSrel}) become
\bea\label{rel-gen-ansatz}
f_1^2(\xi,R)\xi'(r)^2&=&1\label{rel-gen-ansatz1}\\
f_2^2(\xi,R)\frac{\sin^2 \xi(r)}{r^2}\frac{\left(1+\vert z\vert^2\right)^2}{\left(1+\vert R\vert^2\right)^2}\abs{ \frac{dR}{dz}}^2&=&1\label{rel-gen-ansatz2}\\\
f_3^2(\xi,R)\frac{\sin^2 \xi(r)}{r^2}\frac{\left(1+\vert z\vert^2\right)^2}{\left(1+\vert R\vert^2\right)^2}\abs{ \frac{dR}{dz}}^2&=&1.\label{rel-gen-ansatz3}\
\eea
Note that (\ref{rel-gen-ansatz1})-(\ref{rel-gen-ansatz3}) are in fact very similar to (\ref{BPS}), but there is a crucial difference: in the first case the modification is in terms of spacetime coordinates (dreibeins to be precise), while in the latter, the functions $f_i$ depend on the target space coordinates, that is, in the fields themselves. These relations imply the following constraints on the functions $f_i$
\bea
f_1(\xi,R)&=&f_1(\xi),\\
f_2(\xi,R)&=&f_3(\xi,R).
\eea
In addition, if we assume that $f_2(\xi,R)$ factorizes, i.e. $f_2(\xi,R)=f_2(\xi)F(R,\bar{R})$ we have
\bea
f_2^2(\xi)\frac{\sin^2 \xi(r)}{r^2}&=&C,\label{ang1} \\
F(R,\bar{R})\frac{\left(1+\vert z\vert^2\right)^2}{\left(1+\vert R\vert^2\right)^2}\abs{ \frac{dR}{dz}}^2&=&\frac{1}{C}\label{ang2},
\eea
where $C$ is a constant that we will choose to one for simplicity. Let us analyze first the angular equation (\ref{ang2}). A judicious choice  $F$ automatically determines $R$ and define the topological charge of the solution. For example
\bea
F(R,\bar{R})=1&\Rightarrow& R(z)=z,\\
F(R,\bar{R})=\frac{(1+\vert R\vert^2)^2}{4 \vert R\vert (1+\vert R\vert)^2}&\Rightarrow& R(z)=z^2,
\eea
which correspond respectively to a charge 1 and charge 2 BPS solutions. Explicit expressions for $F(R, \bar{R})$ of  topological charges 3 and 4 are not specially illuminating, since the involve inversions of 3 and 4 degree polynomials. For even higher topological charges an explicit expression is not guaranteed. In general, one can compute the corresponding $F(R,\bar{R})$ of any topological charge and with a prescribed symmetry in the following way. First, we choose a rational map  $R=S(z)$ of any degree and invert the equation, $z=S^{-1}(R)$. Then, we choose $F(R,\bar{R})$ such that (\ref{ang2}) is satisfied. The obvious choice is
\be
F(R,\bar{R})=\frac{(1+R\bar{R})^2}{(1+S^{-1}(R) S^{-1}(\bar{R}))^2}\vert R'(S^{-1}(R))\vert^{-2}.
\ee
Note that in this way, $F(R,\bar{R})$ is entirely written in terms of target space variables and satisfies (\ref{ang2}). It still remains to determine the functions $f_1(\xi)$ and $f_2(\xi)$. In principle we can choose almost freely one of the functions, let us say $f_1(\xi)$. Then, by consistency, $f_2(\xi)$ must be chosen in such a way that (\ref{ang1}) is verified by the solution of (\ref{rel-gen-ansatz1}). The consistency condition (remember that we choose $C=1$) between both functions may be written symbolically as follows
\be\label{constraint-f}
\int f_1(\xi)d\xi=f_2(\xi)\sin\xi.
\ee

For example, we can reproduce the solution of Sec. \ref{curved-space} with the choice
\be\label{choice-adam}
f_1^2(\xi)=\frac{1}{(\cos \xi-1)^2}\Rightarrow f_2(\xi)=f_1(\xi)
\ee
Actually, once a choice of the profile functions has been made, it generates a solution in any BPS sector provided that we choose properly the angular function $F(R,\bar{R})$. It other words, it is the choice of the function $F(R,\bar{R})$ that determines which topological sector is BPS. It is clear that all the solutions constructed in this way saturate the energy bound (\ref{BPSrel}) and it is straightforward (but tedious) to prove that the also verify the Euler-Lagrange equations from (\ref{energy-target}). It must be emphasized that the choice (\ref{choice-adam}) is far from unique. If $f_1(\xi)$ is chosen in such a way that (\ref{rel-gen-ansatz1}) satisfies the correct boundary conditions, then $f_2(\xi)$ can be computed formally from (\ref{constraint-f}).



\section{The gravitating Skyrme model}
\label{gravitating}

As we have seen, for a fixed spacetime manifold the Skyrme model admits one BPS sector of arbitrary degree. Let us consider now the coupling to gravity. In this case we have
\be\label{Skyrmegrav}
\mathcal{L}=\frac{1}{2\kappa}\int \sqrt{-g}\left(\mathcal{R}-2\Lambda\right)+\int \sqrt{-g}\left(\frac{f^2}{2}\mathcal{L}_2+\frac{1}{16e^2}\mathcal{L}_4\right),
\ee
where $\Lambda$ is a cosmological constant. We will consider the following metric ansatz
\be
ds^2=-dt^2+g_{ij}dx^i dx^j,
\ee
where $g_{ij}$ is the base space metric and $x^i=(r,\theta,\phi)$. The solutions of the form (\ref{ansatz}) with $\xi(r)$ defined by (\ref{solBPS}) still saturate the energy bound (\ref{bound}) for a metric of the form (\ref{metric}). But now, due to the coupling to gravity, the metric is not fixed, but it must satisfy the Einstein equations
\be\label{Einstein}
G_{\mu\nu}+\Lambda g_{\mu\nu}=\kappa T_{\mu\nu},
\ee
where $G_{\mu\nu}$ is the Einstein tensor and $T_{\mu\nu}$ is the energy-momentum tensor. Let us consider then, the family of 3-manifolds, $\mathcal{M}_3(Q)$  with metric (\ref{metric}) (labeled by the rational function $Q$). The volume and the curvature scalar can be computed easily
\be
\text{Vol}(\mathcal{M}_3(Q))=2\pi r_0^2 \text{deg}Q,\quad \mathcal{R}(\mathcal{M}_3(Q))=\frac{6}{r_0^2}.
\ee  
Now, if we assume that a metric of the form (\ref{metric}) is a solution of the Einstein equations then, the Einstein tensor has a simple form
\bea\label{Eten1}
G_0^{\, 0}&=&-\frac{3}{r_0^2}, \\
G_i^{\, i}&=&-\frac{1}{r_0^2}.\label{Eten2}
\eea
From Sec. \ref{curved-space}, we know that the BPS solutions in the metric (\ref{metric}) satisfy (\ref{eqBPS1}) and (\ref{eqBPS2}), provided that they are given by the rational map ansatz. The energy-momentum tensor in the rational map ansatz can be expanded as follows 
\bea
T_t^{\,t}&=& \Delta_1 -\xi'^2(r) \Delta_2, \\
T_r^{\,r}&=& \Delta_1 +\xi'^2(r) \Delta_2, \\
T_\theta^{\,\theta}&=&\frac{(1+r^2)^2}{16 r_0^4}\left( \frac{(1+r^2)^2\sin^4\xi(r)}{e^2 r^4} -4 f^2 r_0^2\xi'^2(r)\right), \\
T_\phi^{\, \phi}&=& T_\theta^{\,\theta},
\eea
where
\bea
\Delta_1&=&\frac{(1+r^2)^2}{32 e^4 r^4 r_0^4}\sin^2\xi(r)\left((1+r^2)^2\cos 2 \xi(r)-1-r^2(2+r^2+16 e^2f^2f_0^2)\right), \\
\Delta_2&=&\frac{(1+r^2)^2}{8 e^2 r^2 r_0^4})\left((1+r^2)^2\sin^2 \xi(r)+2 e^2 f^2 r^2 r_0^2\right).
\eea
It should be noted that the energy-momentum tensor does not depend on the rational function $R$, but only on the profile function. Our hypothesis is that the BPS solutions already found for a fixed background metric (\ref{solBPS}) are still solutions in the gravitating case. Therefore, one should restrict the energy-momentum tensor to the BPS sector. The on-shell energy-momentum tensor has the form
\bea\label{emBPS1}
T_0^{\, 0}&=&-3\frac{1+ f^2 e^2 r_0^4}{e^2 r_0^4}, \\
T_i^{\, i}&=&-\frac{1-e^2 f^2 r_0^2}{e^2r_0^4}.\label{emBPS2}
\eea
Finally, taking into account (\ref{Eten1})- (\ref{Eten2}) and (\ref{emBPS1})-(\ref{emBPS2}), Einstein equations (\ref{Einstein}) are identically satisfied if the following constraints for the radius of the base space and the cosmological constant hold
\be\label{consistency}
r_0^2=\frac{2\kappa}{e^2(1-\kappa f^2)},\quad \Lambda=\frac{3}{4\kappa }e^2\left(\kappa f^2-1\right)^2.
\ee
This is the consistency condition we were looking for. As long as (\ref{consistency}) holds, field equations and Einstein equations are satisfied simultaneously for a rational function $R$ of any degree. Similar constraints were obtained for the $B=1$ case in \cite{Canfora1,Canfora2}. Moreover, since the condition $ef=1/r_0$ must be met for the BPS equations to be satisfied we have
\be\label{constraints}
\Lambda=e^2 f^2,\quad \kappa=\frac{1}{3 f^2}.
\ee
On constant time hypersurfaces, the topological charge of a BPS soliton is well-defined and determined by the degree of $R$ provided that (\ref{constraints}) are satisfied. Therefore, the model (\ref{Skyrmegrav}) has solutions of arbitrary degree consistent with Einstein equations. It should be noted that for any degree, the radial profile is given by (\ref{solBPS}) with an angular dependence encoded in the rational map ansatz. It is worth noting that the cosmological constant cannot vanish due to the constraints (\ref{constraints}). A vanishing $\Lambda$ implies either a vanishing Dirichlet term or a divergent Skyrme term. Moreover, the base space manifold  must be compact ($r_0$ finite) since the condition $ef=1/r_0$ must hold in order to have BPS solutions. In addition, the size of the Skyrmion $L=r_0=1/\sqrt{\Lambda}$ has to cover the whole base space in order the solution to be BPS. This feature was also observed in \cite{Manton2} for the $B=1$ Skyrmion in $\mathbb{S}^3$ and in \cite{Gudnason} for the sextic (BPS) Skyrme model. 





\section{Conclusions}
\label{conclusions}

We have discussed two modifications of the Skyrme model allowing for self-dual sectors. First, we have proposed a family of curved space manifolds labeled by rational maps. A particular choice of such a map automatically fixes one (and only one) BPS sector of topological charge $N$, being $N$ the degree of the rational map. Moreover, for all possible choices of rational maps, the radial part of the solution takes the same form, and corresponds to the identity map after a suitable change of coordinates. The angular part of the solutions corresponds exactly to the rational map chosen to build  the space metric in the first place. 

In the second part of the paper we have discussed a different deformation of the Skyrme model. This time, instead of curving the space manifold we have added a set of dielectric functions. This could be interpreted as a sort of modification of the target space metric. We have demonstrated that, a particular choice of these field dependent functions brings one topological sector to self-duality. We have also provided an explicit example, based on the rational map ansatz, where the BPS sector in exactly solvable. 

Regarding possible applications to nuclear physics, one should look for near-BPS models rather than strict BPS models. In our case, it is not very difficult to convince oneself that small deformations of the dielectric functions will move the corresponding sector from BPS to near-BPS. A related question concerns the binding energies of those sectors that were not BPS in the original model. The huge freedom in the choice of $F$ and $f_i$ makes very difficult to answer this question in general. At least, for one particular choice \cite{Adam1} the binding energies can be reduced to a $0.2 \% $. Whether or not it is possible to obtain realistic binding energies for other choices of the dielectric functions for a range of baryonic charges will depend of course on the details of the dielectric functions, by we can say \textit{a priori} that, within this family of models, at least one BPS sector (of any charge) will have arbitrarily small binding energy, depending on how the near-BPS deformation is done. On the other hand, a physical motivation for the dielectric functions, whether they are obtained from higher-order terms in the effective field theory perspective or as in-medium coupling constants, may reduce this freedom. We think that these issues deserve further investigation.

It should be noted that the models provided in Secs. \ref{curved-space} and \ref{dielectric} have only one self-dual sector, i.e. they do not correspond to a standard BPS model (where infinite topological sectors give rise to infinite self-dual sectors). However, the gravitating case discussed in Sec. \ref{gravitating} seems to have this property. The Einstein equations impose some constraints between the size of the spacetime manifold and the cosmological constant, or more precisely a relation between both of them and the couplings of the model. In this situation Einstein equations are satisfied for a metric of the form (\ref{metric}) and the rational map ansatz automatically satisfies the BPS equations in all sectors.

Another interesting feature that deserves further study is the possible existence of a moduli space.  As we have mentioned, in the non gravitating case, a choice of rational map $R(z)$ for the space metric, determines the BPS sector and the exact solution in the BPS sector is defined by the rational map itself. This means that, once $R(z)$ has been chosen, say of degree $N$, any other rational map of the same (or different degree) is not a BPS solution. This implies in particular that there is no moduli space associated to that BPS sector. The situation in the gravitating case is different. In Sec. \ref{gravitating} we have shown that, once a cosmological constant has been added, the gravitating model has infinitely many BPS sectors (of any topological charge) whose BPS solutions are given by rational maps and metric solutions given by (\ref{metric}). Since the BPS sector is determined by the degree of the rational map, a trajectory on the moduli space corresponds to the variation of the coefficients of the rational map (preserving the degree), which in turn, implies a variation of the metric of the base space. A detailed analysis of these moduli spaces is left for a future work.

Finally, it would be also interesting to explore the possible supersymmetric structure of the gravitating model presented here. In three spacetime dimensions there are BPS Skyrme models, that is, Skyrme models with infinite self-dual sectors. It is well-known that there is a relation between self-duality and SUSY, and in fact, SUSY versions of these low dimensional models have been found \cite{Queiruga1, Queiruga2, Queiruga3, Bolognesi} some years ago. So far, the attempts of supersymetrization of the Skyrme model in four dimensions were only partial \cite{Nepomechie, Queiruga4, Gudnason1}. Actually, the supersymmetric form of the Skyrme term alone, both in two and three dimensions, requires the introduction of a constant (\cite{Queiruga2, Gudnason1}) that depends on the couplings as (\ref{constraints}), but it does not allow for a Dirichlet term (although in two dimensions this constant can be promoted to a field dependent potential preserving SUSY). The gravitating Skyrme model described in Sec. \ref{gravitating}  suggests that, there should be a SUGRA formulation of this model. These problems are also left for a future investigation.


{\bf Acknowledgements.}- This work is supported by the Spanish Ministry MCIU/AEI/FEDER grant (PGC2018-094626-B-C21), the Basque Government grant (IT-979-16). The author is grateful to C. Adam and A. Wereszczynski for useful comments.

 
\renewcommand{\theequation}{A.\arabic{equation}}
\setcounter{equation}{0}



\end{document}